\documentclass[12pt]{article}

\usepackage{graphicx}

\renewcommand{\thefootnote}{\fnsymbol{footnote}}

\begin{document}

\vspace*{-0.8cm}
\begin{flushright}
DPNU-05-04
\end{flushright}
\vspace{-0.8cm}
\begin{center}
\Large\bf
Hidden LocaL Symmetry Theory\\
as an Effective Field Theory of QCD
\footnote{%
Talk given at International workshop on
``Dynamical Symmetry Breaking''
(December 21-22, 2004, Nagoya, Japan).}
\end{center}

\begin{center}
{\large Masayasu Harada}
\end{center}

\begin{center}
{\it Department of Physics, Nagoya University, Nagoya 464-8602, Japan}
\end{center}

\begin{abstract}
In this write-up, 
I summarize the recent development of the hidden local symmetry 
(HLS) theory as an effective field theory of QCD.
I first explain how the 
systematic chiral perturbation in the HLS
is justified in the large $N_c$ limit of QCD, and
summarize the
basic concept of the Wilsonian matching,
through which some of the bare parameters of the HLS are 
determined by matching the HLS to the operator product
expansion in QCD.
Then, I briefly review how to formulate
the vector manifestation in hot matter.
\end{abstract}

\setcounter{footnote}{0}
\renewcommand{\thefootnote}{\#\arabic{footnote}}

\section{Introduction}

Quantum Chromodynamics (QCD)
is known to be a fundamental theory for describing the
low-energy hadron phenomena.
However, it is very difficult to reproduce experimental data
directly from QCD,
since QCD is the strong coupling gauge theory.
Then,
instead of studying QCD directly,
it is convenient to 
use effective
field theories (EFTs) written in terms of hadronic degrees of
freedom.
When one studies the phenomena of light hadrons,
it is important for EFTs to reproduce the chiral symmetry
properties of QCD:
The Lagrangian of QCD in the light quark sector possesses 
the approximate chiral symmetry which is spontaneously broken
down to the flavor symmetry.  As a result,
the pion appears as the approximate massless Nambu-Goldstone
boson.
Furthermore, 
there must exist a systematic expansion scheme in an EFT,
by which one can systematically 
include higher derivative terms together with 
loop corrections.

One of popular EFTs is the chiral perturbation theory 
(ChPT)~\cite{ChPT:Weinberg,GL}, in which
the pion is the only relevant degree of freedom.
Starting from the energy region around the pion mass,
one can study the higher energy region systematically
by including the higher derivative terms together with
the loop corrections.
In much higher region, 
however, we know the existence of $\rho$ meson 
and the ChPT may not be applicable in
the energy scale beyond the $\rho$ meson mass.
One simple way to study the hadron phenomena in such
an energy region
is to include $\rho$ meson in addition to the pion 
and make an EFT.

There are several ways to include the rho meson into effective
Lagrangians in the literature:
the matter field method~\cite{CWZ:CCWZ};
the massive Yang-Mills field method~\cite{massiveYM};
the anti-symmetric tensor field method~\cite{GL,Eck:89a};
and the hidden local symmetry (HLS)~\cite{BKUYY:BKY}.
In the HLS,
as first pointed 
in Ref.~\cite{Georgi} and
developed further in Refs.~\cite{HY:PLB,Tanabashi,HY:WM,HY:PR},
thanks to the gauge symmetry,
a systematic loop expansion can be performed with
the $\rho$ meson included in addition to the pion.
Furthermore, in Ref.~\cite{HY:WM}, 
the Wilsonian matching was
proposed as a way to determine some of the
parameters by matching the HLS to QCD, which 
together with the ChPT with HLS
was done in remarkable agreement with the 
experiments~\cite{HY:WM,HY:PR,FHS}.

In Ref.~\cite{HY:VM},
using the Wilsonian matching, we studied
the chiral symmetry restoration in large flavor 
QCD~\cite{rest:Nf}, 
and proposed
a new pattern of the chiral symmetry restoration,
which we called 
the vector manifestation (VM).
The VM is a novel manifestation of Wigner 
realization of
chiral symmetry where the $\rho$ meson becomes massless 
degenerate with the pion at the
chiral phase transition point.
In Refs.~\cite{HS:VMT,HKR},
the Wilsonian matching together with the HLS was applied 
for the chiral symmetry restoration in hot and/or dense 
QCD~\cite{rest1,Brown-Rho:96,Rapp-Wambach:00},
and the formulations of the VM in hot matter~\cite{HS:VMT}
and dense matter~\cite{HKR} were presented.
The VM in hot and/or dense matter
gives a theoretical support of the dropping mass of the 
$\rho$ meson
following the Brown-Rho scaling proposed in Ref.~\cite{BR},
which can explain 
(see, e.g., Refs.~\cite{Li:1995qm, Brown-Rho:96, Rapp-Wambach:00})
the enhancement of dielectron ($e^+e^-$) mass spectra
below the $\rho / \omega$ resonance
observed at the 
CERN Super Proton Synchrotron (SPS)~\cite{Agakishiev:1995xb}.

In this write-up,
I first summarize the recent development of the HLS theory
as an EFT of QCD including the Wilsonian matching.
Then, I will briefly show how to formulate the VM in hot matter.

This write-up is organized as follows:
In section~\ref{sec:HLS},
starting from the Lagrangian of the HLS,
I will briefly explain how the ChPT with HLS is justified
in the large $N_c$ limit of QCD.
Next, I will 
introduce several essential ingredients of the 
Wilsonian matching in section~\ref{sec:WM}.
In section~\ref{sec:VM}, I will 
briefly review the difference between the 
VM and
the conventional manifestation of chiral symmetry restoration
based on the linear sigma model.
Section~\ref{sec:VMT} is devoted to show
how to formulate the VM in hot matter.
Finally, in section~\ref{sec:sum},
I will give a brief summary.

\section{Hidden Local Symmetry}
\label{sec:HLS}

In this section I will
briefly explain the hidden local symmetry (HLS) theory.

The HLS theory is based on 
the $G_{\rm{global}} \times H_{\rm{local}}$ symmetry,
where $G=SU(N_f)_{\rm L} \times SU(N_f)_{\rm R}$ 
is the chiral symmetry
and $H=SU(N_f)_V$ is the HLS. ~\footnote{
 In this write-up, I consider the QCD with general number of flavors,
 i.e., I include $N_f$ massless quarks.  Nevertheless, I call the
 vector meson ($\rho$ meson and its flavor partner) the $\rho$, and the
 pseudoscalar meson (pion and its flavor partner) the $\pi$.
}
The basic quantities are 
the HLS gauge boson and two matrix valued
variables $\xi_{\rm L}(x)$ and $\xi_{\rm R}(x)$
which transform as
 \begin{equation}
  \xi_{\rm L,R}(x) \to \xi^{\prime}_{\rm L,R}(x)
  =h(x)\xi_{\rm L,R}(x)g^{\dagger}_{\rm L,R}\ ,
 \end{equation}
where $h(x)\in H_{\rm{local}}$ and $g_{\rm L,R}\in
[\mbox{SU}(N_f)_{\rm L,R}]_{\rm{global}}$.
These variables are parameterized as
 \begin{equation}
  \xi_{\rm L,R}(x)=e^{i\sigma (x)/{F_\sigma}}
   e^{\mp i\pi (x)/{F_\pi}}\ ,
 \end{equation}
where $\pi = \pi^a T_a$ denotes the pseudoscalar 
Nambu-Goldstone (NG) bosons
associated with the spontaneous symmetry breaking of
$G_{\rm{global}}$ chiral symmetry, 
and $\sigma = \sigma^a T_a$ denotes
the NG bosons associated with 
the spontaneous breaking of $H_{\rm{local}}$.
This $\sigma$ is absorbed into the HLS gauge 
boson through the Higgs mechanism. 
$F_\pi$ and $F_\sigma$ are the decay constants
of the associated particles.
The phenomenologically important parameter $a$ is defined as 
 \begin{equation}
  a = \frac{{F_\sigma}^2}{{F_\pi}^2}\ .
 \end{equation}
The covariant derivatives of $\xi_{\rm L,R}$ are given by
\begin{eqnarray}
 D_\mu \xi_{\rm L} &=& \partial_\mu\xi_{\rm L} - 
  iV_\mu \xi_{\rm L} + i\xi_{\rm L} {\cal{L}}_\mu\ ,
 \nonumber\\
 D_\mu \xi_{\rm R} &=& \partial_\mu\xi_{\rm R} - 
  iV_\mu \xi_{\rm R} + i\xi_{\rm R} {\cal{R}}_\mu\ ,
\end{eqnarray}
where $V_\mu$ is the gauge field of $H_{\rm{local}}$, and
${\cal{L}}_\mu$ and ${\cal{R}}_\mu$ are the external
gauge fields introduced by gauging the 
$G_{\rm{global}}$ symmetry.

The HLS Lagrangian with the lowest derivative terms in the 
chiral limit is given by~\cite{BKUYY:BKY}
 \begin{equation}
  {\cal{L}}_{(2)} = {F_\pi}^2\mbox{tr}\bigl[ \hat{\alpha}_{\perp\mu}
                                      \hat{\alpha}_{\perp}^{\mu}
                                   \bigr] +
       {F_\sigma}^2\mbox{tr}\bigl[ \hat{\alpha}_{\parallel\mu}
                  \hat{\alpha}_{\parallel}^{\mu}
                  \bigr] -
        \frac{1}{2g^2}\mbox{tr}\bigl[ V_{\mu\nu}V^{\mu\nu}
                   \bigr]
\ , \label{eq:L(2)}
 \end{equation}
where $g$ is the HLS gauge coupling,
$V_{\mu\nu}$ is the field strength
of $V_\mu$ and
 \begin{eqnarray}
  \hat{\alpha}_{\perp }^{\mu}
     &=& \frac{1}{2i}\bigl[ 
        D^\mu\xi_{\rm R} \cdot \xi_{\rm R}^{\dagger} -
        D^\mu\xi_{\rm L} \cdot \xi_{\rm L}^{\dagger}
                   \bigr] \ ,
\nonumber\\
  \hat{\alpha}_{\parallel}^{\mu}
     &=& \frac{1}{2i}\bigl[ 
          D^\mu\xi_{\rm R} \cdot \xi_{\rm R}^{\dagger}+
          D^\mu\xi_{\rm L} \cdot \xi_{\rm L}^{\dagger}
                   \bigr]
\ .
 \end{eqnarray}

In the HLS,
as first pointed 
in Ref.~\cite{Georgi} and
developed further in Refs.~\cite{HY:PLB,Tanabashi,HY:WM,HY:PR},
thanks to the gauge symmetry,
a systematic loop expansion can be performed with
the vector mesons included in addition to the pseudoscalar mesons.
In this chiral perturbation theory (ChPT) with  HLS 
one can show that the loop expansion corresponds
to the derivative expansion
as in the ordinary ChPT.
Here, I show the expansion parameter and the order counting
of the systematic expansion in the HLS.
As is well known, the expansion parameters of the ordinary
ChPT in the chiral limit
is $p^2/\Lambda_\chi^2$,
where $p$ is the typical momentum scale and
$\Lambda_\chi \sim 4 \pi F_\pi$ the chiral symmetry breaking scale.
In addition to these two parameters,
$m_\rho^2/\Lambda_\chi^2$ is also the expansion parameter in the HLS.
For the validity of the expansion in the parameter
$m_\rho^2/\Lambda_\chi^2$,
we need to show that 
this expansion parameter is actually small, and that
the quantum correction proportional to $1/m_\rho$
never appears.

Let me first consider the smallness of the expansion 
parameter $m_\rho^2/\Lambda_\chi^2$.
The smallness is actually justified in the large $N_c$ QCD.
As is well known, in the large $N_c$ limit of QCD,
the $\pi$ decay constant scales as $F_\pi \sim \sqrt{N_c}$,
while the $\rho$ mass does not scale.
So the ratio $m_\rho^2/(4\pi F_\pi)^2$ scales as $1/N_c$, and
becomes small in the large $N_c$ limit.
In this way, the smallness of the
expansion parameter $m_\rho^2/\Lambda_\chi^2$
is justified in the large $N_c$ QCD:
\begin{equation}
\frac{m_\rho^2}{\Lambda_\chi^2} = \frac{m_\rho^2}{(4\pi F_\pi)^2}
 \sim \frac{1}{N_c} \ \ll \ 1 \ .
\end{equation}
One should note that
this argument is true for any models including $\rho$,
which is not enough for the existence of a systematic
expansion.

For the existence of a systematic expansion,
one needs to show that the contribution proportional to $1/m_\rho$
never appears at any loop order.
This is actually
guaranteed by the gauge invariance
in the HLS theory, while there is no such argument in other 
models in my best knowledge\footnote{%
 For recent attempts to include the effects of dynamical $\rho$
 in models other than the HLS, see, e.g., Ref.~\cite{vector1}.
}.
For example,
when we include $\rho$ as the matter field in the
chiral Lagrangian, the form of the propagator of $\rho$ is
given by 
\begin{equation}
\frac{1}{p^2-m_\rho^2} 
\left[ g_{\mu\nu} - \frac{p_\mu p_\nu}{m_\rho^2} \right] \ ,
\label{rho:prop0}
\end{equation}
which coincides with the $\rho$ propagator in the unitary gauge
of the HLS (Weinberg's $\rho$ meson~\cite{Weinberg:68}).
The longitudinal part ($p_\mu p_\nu$-part) carries the factor of
$1/m_\rho^2$ which may generate quantum corrections
proportional to some powers of $1/m_\rho^2$.
Appearance of a factor
$1/m_\rho^2$ is a disaster in the loop calculations, particularly when
the $\rho$ mass is light.
Namely, the derivative expansion discussed above breaks down.
We note that the situation is similar in the ``Massive Yang-Mills''
approach and the ``anti-symmetric tensor field method''.

In the HLS, however, the gauge invariance prevent such a $1/m_\rho^2$
factor from appearing.
This can be easily seen by the following $\rho$ propagator in an
$R_\xi$-like gauge fixing~\cite{HY:PLB}:
\begin{equation}
\frac{1}{p^2-m_\rho^2} 
\left[ 
  g_{\mu\nu} - (1-\alpha) \frac{p_\mu p_\nu}{p^2- \alpha m_\rho^2}
\right] \ ,
\label{rho:prop1}
\end{equation}
where $\alpha$ is the gauge fixing parameter.
The propagator in Eq.~(\ref{rho:prop1}) is well defined in the limit
of $m_\rho\rightarrow0$ except for the unitary gauge 
($\alpha =\infty$), 
while the propagator in
Eq.~(\ref{rho:prop0}) is ill-defined in such a limit.
In addition, the gauge invariance guarantees that all the
interactions never include a factor of $1/g^2 \propto 1/m_\rho^2$,
while it may exist for the lack of the gauge invariance.
Then all the loop corrections are well defined
even in the limit of $m_\rho\rightarrow0$.
Thus the HLS gauge invariance is essential to performing the above
derivative expansion.
This makes the HLS most powerful among various methods 
for including the
vector mesons based on the chiral symmetry.

One may think that 
the scalar mesons should be included,
since several analysis~\cite{scalars}
shows that they are
lighter than the vector mesons in real-life QCD.
For example, the analysis in Ref.~\cite{HSS} shows that
the mass of sigma meson is about 560\,MeV, 
which is definitely lighter than
the $\rho$ meson mass, $m_\rho = 770$\,MeV.
In the large $N_c$ limit of QCD, 
however, it is natural to assume that
the {\it light} sigma meson (flavor singlet scalar meson)
does not exist 
in the following sense:
There are two major pictures on the composition of the sigma meson,
i.e., 2-quark picture and 4-quark picture~\cite{Jaffe,BFSS}.
In the 2-quark picture,
the sigma meson is made of one quark and one anti-quark
and it is much lighter than the $a_0(980)$ meson due to
the instanton effect~\cite{scalarcollab}.
Then, in the 2-quark picture,
one can expect that the mass of the sigma meson becomes
heavier and agree with the mass of $a_0(980)$ meson
in the large $N_c$ limit of QCD.
In the 4-quark picture, on the other hand,
the sigma meson is made of two quarks and
two anti-quarks, which does not exist in the large $N_c$ limit
of QCD~\cite{Jaffe,Witten}.
Furthermore,
recently in Ref.~\cite{HSS:04},
the analysis adopted in Refs.~\cite{SS:95,HSS}
was extended
for studying the $\pi$-$\pi$ scattering in the real-life QCD
to the one in the large $N_c$ QCD, and it was shown
that, for $N_c \ge 6$,
the unitarity in the scalar channel of the $\pi$-$\pi$ scattering 
is satisfied without
scalar mesons up until the energy scale of $4\pi F_\pi$.
This indicates that we do not need scalar mesons in the
low-energy region in the large $N_c$ QCD.
In the real-life QCD
the sigma meson is lighter than the $\rho$ meson, but it
is actually very broad.
I expect that loop corrections from 
such broad resonances are very small, and
that the chiral perturbation in the HLS
is still possible, as far as we do not see the
scalar channel.

Now that I explained that the smallness of the expansion parameter
$m_\rho^2/\Lambda_\chi^2$ is justified in the large $N_c$ QCD,
and that no appearance of quantum corrections
proportional to some powers of $1/m_\rho^2$,
I show the chiral order counting in the ChPT with HLS.
As in the ordinary ChPT,
the derivative and the external gauge fields are counted as
${\cal O}(p)$:
$\partial_\mu \sim {\cal L}_\mu \sim 
  {\cal R}_\mu \sim {\cal O}(p)$.
In the HLS, $\rho$ acquires its mass through the Higgs
mechanism, which implies that the $\rho$ mass $m_\rho$
is proportional to the gauge coupling $g$.
Then, the smallness of $m_\rho/\Lambda_\chi$ is achieved by
the small gauge coupling.
The expansion of the HLS is
done by considering the expansion parameter $m_\rho/\Lambda_\chi$
to be equally small as $p/\Lambda_\chi$.
Thus, the gauge coupling is counted as order 
$p$~\cite{Georgi,Tanabashi}: 
\begin{equation}
 g \sim {\cal O}(p).
\end{equation}
This is the most important part of the ChPT with HLS.
Using the above counting scheme,
one can show that the loop expansion corresponds to the 
low-energy expansion and
systematically calculate 
quantum corrections to several physical
quantities based on the ChPT with HLS.

According to the entire list shown in Ref.~\cite{Tanabashi},
there are 35 counter terms at ${\cal O}(p^4)$ for general
$N_f$.
However, only three terms are relevant in the present analysis
in which I consider two-point functions
in the chiral limit:
 \begin{equation}
  {\cal{L}}_{(4)} = z_1\mbox{tr}\bigl[ \hat{\cal{V}}_{\mu\nu}
                       \hat{\cal{V}}^{\mu\nu} \bigr] +
                    z_2\mbox{tr}\bigl[ \hat{\cal{A}}_{\mu\nu}
                       \hat{\cal{A}}^{\mu\nu} \bigr] +
                    z_3\mbox{tr}\bigl[ \hat{\cal{V}}_{\mu\nu}
                       V^{\mu\nu} \bigr], \label{eq:l(4)}
 \end{equation}
where
 \begin{eqnarray}
  \hat{\cal{A}}_{\mu\nu}=\frac{1}{2}
                   \bigl[ \xi_R{\cal{R}}_{\mu\nu}\xi_R^{\dagger}-
                          \xi_L{\cal{L}}_{\mu\nu}\xi_L^{\dagger}
                   \bigr]\ ,
  \label{def a mn}\\
  \hat{\cal{V}}_{\mu\nu}=\frac{1}{2}
                    \bigl[ \xi_R{\cal{R}}_{\mu\nu}\xi_R^{\dagger}+
                           \xi_L{\cal{L}}_{\mu\nu}\xi_L^{\dagger}
                    \bigr]\ ,
  \label{def v mn}
 \end{eqnarray}
with ${\cal{R}}_{\mu\nu}$ and ${\cal{L}}_{\mu\nu}$ being
the field strengths of 
${\cal{R}}_{\mu}$ and ${\cal{L}}_{\mu}$.

\section{Wilsonian Matching}
\label{sec:WM}

The values of ${\cal O}(p^4)$ parameters 
as well as the leading order parameters $F_\pi$, $a$ and $g$ 
in the ChPT with HLS introduced in the previous section
should be determined from 
the underlying QCD.
In Ref.~\cite{HY:WM}, we proposed a way to determine some of the
parameters by matching the HLS to QCD, which we called the
Wilsonian matching.
In this section, I will briefly review the Wilsonian matching.

Let me first explain the basic concept of the Wilsonian matching.
In the high energy region,
quarks and gluons are good degrees of freedom,
and one can treat QCD in a perturbative way.
In the low energy region, on the other hand,
hadrons become good degrees of freedom instead of 
quarks and gluons.
The main assumption of the Wilsonian matching is that
there is some scale $\Lambda$ at which both the perturbative
QCD and the ChPT with HLS are applicable, and that
one can switch the theory from QCD to the HLS at $\Lambda$.
When there is such an overlapping energy region,
the bare parameters of the HLS
can be determined by matching the HLS
with QCD at the matching scale $\Lambda$.
In other words,
the bare theory of the HLS can be
obtained by integrating out the high energy modes, i.e.,
quarks and gluons, at $\Lambda$.
Once the bare theory is determined,
the quantum effect is included to
relate the bare parameters with physical quantities such as the
$\pi$ decay constant and the $\rho$ mass.

Since the procedure of the Wilsonian matching
is a little different from the one adopted 
in several other EFTs,
I show the essential difference 
starting with 
the basic concept of the EFT: The effective Lagrangian of the EFT,
which has the most general form constructed from the chiral symmetry,
gives the same generating functional as that obtained from QCD:
\begin{equation}
 Z_{\rm EFT}[J,F] = \int {\cal D}U e^{i S_{\rm eff}[J,F]}
 \mathop{\leftrightarrow}_{\rm matching}
 Z_{\rm QCD}[J] = \int {\cal D}q {\cal D}\bar{q} {\cal D}G
         e^{i S_{\rm QCD}[J]}
\ ,
\label{gf-match}
\end{equation}
where $J$ is a set of external source fields.
In the EFT side,
$U$ denotes a set of 
the relevant hadronic fields such as the pion fields,
$S_{\rm eff}$ is the action 
expressed in terms of these hadrons,
and $F$ a set of parameters included in the EFT.
In QCD side,
$q$ $(\bar{q})$ denotes (anti) quark field,
$G$ is the gluon field and $S_{\rm QCD}$ represents the action
expressed in terms of the quarks and gluons.

In some matching schemes, the renormalized parameters of the EFT
are determined by the matching.
On the other hand,
the matching in the Wilsonian sense is performed based on the 
following general idea:
The bare Lagrangian of the EFT is defined at a suitable 
matching scale $\Lambda$ and
the generating functional derived from the bare Lagrangian 
leads to the same Green's function as that derived in QCD
at $\Lambda$:
\begin{equation}
\bigl.
 Z_{\rm EFT}[J,F] 
\bigr\vert_{E = \Lambda}
= e^{i S_{\rm eff}[J,F_{\rm bare}]}
 \ \mathop{\longleftrightarrow}_{\rm matching}\ 
\bigl.
 Z_{\rm QCD}[J] 
\bigr\vert_{E = \Lambda}
= \int {\cal D}q {\cal D}\bar{q} {\cal D}G
         e^{i S_{\rm QCD}[J]}
\ ,
\label{gf-bare-match}
\end{equation}
where $F_{\rm bare}$ denotes a set of bare parameters.
Through the above matching, 
the {\it bare} parameters of the EFT are determined.
In other words,
one obtains the bare Lagrangian of the EFT after
integrating out the high energy modes, i.e.,
the quarks and gluons above $\Lambda$.
Then the informations of the high energy modes are included in the
parameters of the EFT.

In Refs.~\cite{HY:WM,HY:PR}, based on the above idea, 
the vector and axial-vector current correlators derived
from the bare HLS theory are matched with those obtained
by the operator product expansion (OPE) in QCD.
It was shown that the physical predictions are in remarkable
agreement with experiments.
Furthermore, a recent analysis~\cite{FHS} shows that
the Wilsonian matching with the effect of current quark masses
included reproduces the ratio $f_k/f_\pi$ in remarkable
agreement with experiment.
I would like to stress that, for the above success of the
Wilsonian matching, the effect
of quadratic divergence plays an essential role:
For obtaining the physical quantities starting from the
{\it bare} theory,
one, of course, has to include the effect of quadratic
divergence into the RGEs in the Wilsonian sense.

Before going to the analysis of the chiral symmetry restoration
based on the Wilsonian matching,
let me discuss the validity of the systematic expansion 
in the Wilsonian matching procedure.
One might think that the systematic
expansion would break down near the matching scale,
since the quadratic divergences from higher loops can in principle
contribute to the ${\cal O}(p^2)$ terms.
However, 
even when
the quadratic divergences are explicitly included,
the systematic expansion is still valid 
in the following sense:
When one starts from the bare theory and calculates the quantum
corrections including the quadratic divergence,
the loop corrections
are given in terms of the bare parameter $F_{\pi,{\rm bare}}$
instead of the on-shell decay constant $F_\pi$.
Then,
the scale at which the theory breaks down 
should be 
\begin{equation}
\Lambda_\chi \simeq 4 \pi F_{\pi,{\rm bare}}
\ .
\label{fnda lam chi 1}
\end{equation}
By using this chiral symmetry breaking scale,
the 
quadratically divergent correction to the ${\cal O}(p^2)$ term 
at $n$th loop order takes the form of
$[\Lambda^2/\Lambda_\chi^2]^{n}$.
As for the ChPT with HLS explained in the previous section,
the requirement
$\Lambda < \Lambda_\chi$ is satisfied in 
the large $N_c$ limit of QCD:
In the large $N_c$ limit of QCD,
the quadratically divergent correction at $n$th loop order
is suppressed by 
$[\Lambda^2/\Lambda_\chi^2]^n \sim [1/N_c]^n$.
As a result,
one can perform the systematic loop expansion with
quadratic divergences included in the large $N_c$ limit,
and extrapolate the results to the real-life QCD.
The quantitative analyses was done in 
Refs.~\cite{HY:WM,HY:PR,FHS}, which show that
the phenomenological analysis based on the Wilsonian matching
together with the ChPT with HLS
can be done in remarkable agreement with the experiments
in much the same sense as the phenomenological analysis
in the ordinary ChPT
is successfully extended to the energy 
region higher than the pion mass scale, which is 
logically beyond the
validity region of the ChPT.

\section{Vector Manifestation of Chiral Symmetry}
\label{sec:VM}

In this section, following Ref.~\cite{HY:VM,HY:PR},
I briefly review the difference between the 
vector manifestation (VM) and
the conventional manifestation of chiral symmetry restoration
based on the linear sigma model
in terms of the chiral representation of the mesons
by extending the analyses done in
Refs.~\cite{Gilman-Harari,Weinberg:69}
for two flavor QCD.

The VM was first proposed in
Ref.~\cite{HY:VM} as a novel manifestation of Wigner 
realization of
chiral symmetry where the vector meson $\rho$ becomes massless at the
chiral phase transition point. 
Accordingly, the (longitudinal) $\rho$ becomes the chiral partner of
the Nambu-Goldstone (NG) boson $\pi$.
The VM is characterized by
\begin{equation}
\mbox{(VM)} \qquad
f_\pi^2 \rightarrow 0 \ , \quad
m_\rho^2 \rightarrow m_\pi^2 = 0 \ , \quad
f_\rho^2 / f_\pi^2 \rightarrow 1 \ ,
\label{VM def}
\end{equation}
where $f_\rho$ is the decay constant of 
(longitudinal) $\rho$ at $\rho$ on-shell.
This is completely different from 
the conventional picture based
on the linear sigma model 
where the scalar meson $S$ becomes massless
degenerate with $\pi$ as the chiral partner:
\begin{equation}
\mbox{(GL)} \qquad
f_\pi^2 \rightarrow 0 \ , \quad
m_S^2 \rightarrow m_\pi^2 = 0 \ .
\label{GL def}
\end{equation}
In Ref.~\cite{HY:PR}
this was called GL manifestation after the
effective theory of Ginzburg--Landau or Gell-Mann--Levy.

I first consider 
the representations of 
the following zero helicity ($\lambda=0$) states
under
$\mbox{SU(3)}_{\rm L}\times\mbox{SU(3)}_{\rm R}$;
the $\pi$, the (longitudinal) $\rho$, the (longitudinal) axial-vector
meson denoted by $A_1$ ($a_1$ meson and its flavor partners)
and the scalar meson denoted by $S$.
The $\pi$ and the longitudinal $A_1$ 
are admixture of $(8\,,\,1) \oplus(1\,,\,8)$ and 
$(3\,,\,3^*)\oplus(3^*\,,\,3)$
since the symmetry is spontaneously
broken~\cite{Gilman-Harari,Weinberg:69}:
\begin{eqnarray}
\vert \pi\rangle &=&
\vert (3\,,\,3^*)\oplus (3^*\,,\,3) \rangle \sin\psi
+
\vert(8\,,\,1)\oplus (1\,,\,8)\rangle  \cos\psi
\ ,
\nonumber
\\
\vert A_1(\lambda=0)\rangle &=&
\vert (3\,,\,3^*)\oplus (3^*\,,\,3) \rangle \cos\psi 
- \vert(8\,,\,1)\oplus (1\,,\,8)\rangle  \sin\psi
\ ,
\label{mix pi a}
\end{eqnarray}
where the experimental value of the mixing angle $\psi$ is 
given by approximately 
$\psi=\pi/4$~\cite{Gilman-Harari,Weinberg:69}.  
On the other hand, the longitudinal $\rho$
belongs to pure $(8\,,\,1)\oplus (1\,,\,8)$
and the scalar meson to 
pure $(3\,,\,3^*)\oplus (3^*\,,\,3)$:
\begin{eqnarray}
\vert \rho(\lambda=0)\rangle &=&
\vert(8\,,\,1)\oplus (1\,,\,8)\rangle  
\ ,
\nonumber
\\
\vert S\rangle &=&
\vert (3\,,\,3^*)\oplus (3^*\,,\,3) \rangle 
\ .
\label{rhos}
\end{eqnarray}

When the chiral symmetry is restored at the
phase transition point, 
it is natural to expect that
the chiral representations coincide with the mass eigenstates:
The representation mixing is dissolved.
{}From Eq.~(\ref{mix pi a}) one can easily see
that
there are two ways to express the representations in the
Wigner phase of the chiral symmetry:
The conventional GL manifestation
corresponds to 
the limit $\psi \rightarrow \pi/2$ in which
$\pi$ is in the representation
of pure $(3\,,\,3^*)\oplus(3^*\,,\,3)$ 
together with the scalar meson, 
both being the chiral partners:
\begin{eqnarray}
\mbox{(GL)}
\qquad
\left\{
\begin{array}{rcl}
\vert \pi\rangle\,, \vert S\rangle
 &\rightarrow& 
\vert  (3\,,\,3^\ast)\oplus(3^\ast\,,\,3)\rangle\ ,
\\
\vert \rho (\lambda=0) \rangle \,,
\vert A_1(\lambda=0)\rangle  &\rightarrow&
\vert(8\,,\,1) \oplus (1\,,\,8)\rangle\ .
\end{array}\right.
\end{eqnarray}
On the other hand, the VM corresponds 
to the limit $\psi\rightarrow 0$ in which the $A_1$ 
goes to a pure 
$(3\,,\,3^*)\oplus (3^*\,,\,3)$, now degenerate with
the scalar meson $S$ in the same representation, 
but not with $\rho$ in 
$(8\,,\,1)\oplus (1\,,\,8)$:
\begin{eqnarray}
\mbox{(VM)}
\qquad
\left\{
\begin{array}{rcl}
\vert \pi\rangle\,, \vert \rho (\lambda=0) \rangle
 &\rightarrow& 
\vert(8\,,\,1) \oplus (1\,,\,8)\rangle\ ,
\\
\vert A_1(\lambda=0)\rangle\,, \vert s\rangle  &\rightarrow&
\vert  (3\,,\,3^\ast)\oplus(3^\ast\,,\,3)\rangle\ .
\end{array}\right.
\end{eqnarray}
Namely, the
degenerate massless $\pi$ and (longitudinal) $\rho$ at the 
phase transition point are
the chiral partners in the
representation of $(8\,,\,1)\oplus (1\,,\,8)$.

Next, I consider the helicity $\lambda=\pm1$. 
note that
the transverse $\rho$
can belong to the representation different from the one
for the longitudinal $\rho$ ($\lambda=0$) and thus can have the
different chiral partners.
According to the analysis in Ref.~\cite{Gilman-Harari},
the transverse components of $\rho$ ($\lambda=\pm1$)
in the broken phase
belong to almost pure
$(3^*\,,\,3)$ ($\lambda=+1$) and $(3\,,\,3^*)$ ($\lambda=-1$)
with tiny mixing with
$(8\,,\,1)\oplus(1\,,\,8)$.
Then, it is natural to consider in VM that
they become pure $(3\,,\,3^\ast)$ and 
$(3^\ast\,,\,3)$
in the limit approaching the chiral restoration point~\cite{HY:PR}:
\begin{eqnarray}
\vert \rho(\lambda=+1)\rangle \rightarrow 
  \vert (3^*,3)\rangle\ ,\quad
\vert \rho(\lambda=-1)\rangle \rightarrow 
  \vert (3,3^*)\rangle \ .
\end{eqnarray}
As a result,
the chiral partners of the transverse components of $\rho$ 
in the VM
will be  themselves. 
Near the critical point the longitudinal $\rho$ becomes 
almost $\sigma$, namely the would-be NG boson $\sigma$ almost 
becomes a 
true NG boson and hence a different particle than the transverse
$\rho$.

\section{Formulation of the Vector Manifestation in Hot Matter}
\label{sec:VMT}

In this section
I briefly review how to formulate the vector manifestation
(VM) in hot matter.
I first show how to
extend the Wilsonian matching
to the version at non-zero temperature in order to incorporate
the intrinsic thermal effect into the bare parameters of the HLS
Lagrangian. 
Then, I briefly summarize
how the VM is formulated in
hot matter following Refs.~\cite{HS:VMT,HS:VD}.
It should be noticed that
the critical temperature of the
chiral symmetry restoration is approached 
from the broken phase up to $T_c - \epsilon$,
and that
the following basic assumptions are adopted in the present
analysis:
(1) The relevant degrees of freedom until near $T_c - \epsilon$
are only $\pi$ and $\rho$;
(2) Other mesons such as $a_1$ and sigma mesons are still heavy 
at $T_c - \epsilon$;
(3) Partial chiral symmetry restoration already occurs at 
$T_c - \epsilon$.
Based on these assumptions, I will show that the VM 
necessarily occurs at the chiral symmetry restoration point.

Let me first explain how to extend
the Wilsonian matching proposed at 
$T=0$~\cite{HY:WM} (see section~\ref{sec:WM})
to the one at non-zero temperature following Ref.~\cite{HS:VD}.
For this, it should be noticed that
there is no longer Lorentz symmetry
in hot matter,
and
the Lorentz non-scalar operators such as
$\bar{q}\gamma_\mu D_\nu q$ may exist in 
the form of the current correlators derived by the 
OPE~\cite{HKL,FLK}.
This leads to, e.g., a difference between the temporal and spatial
bare $\pi$ decay constants.
In the present analysis,
however, I neglect the contributions from these operators
since they give a small correction compared with 
the main term $1 + \frac{\alpha_s}{\pi}$.
This implies that the Lorentz symmetry breaking effect in
the bare $\pi$ decay constant is small, 
$F_{\pi,\rm{bare}}^t \simeq F_{\pi,\rm{bare}}^s$~\cite{HKRS}.
Thus it is a good approximation that I determine the $\pi$ decay
constant at non-zero temperature through the matching 
condition obtained at $T=0$ in Ref.~\cite{HY:WM}
with putting possible temperature dependences on the gluonic 
and quark condensates~\cite{HS:VMT,HKRS}:
\begin{equation}
 \frac{F^2_\pi (\Lambda ;T)}{{\Lambda}^2} 
  = \frac{1}{8{\pi}^2}\Bigl[ 1 + \frac{\alpha _s}{\pi} +
     \frac{2{\pi}^2}{3}\frac{\langle \frac{\alpha _s}{\pi}
      G_{\mu \nu}G^{\mu \nu} \rangle_T }{{\Lambda}^4} +
     {\pi}^3 \frac{1408}{27}\frac{\alpha _s{\langle \bar{q}q
      \rangle }^2_T}{{\Lambda}^6} \Bigr]
\ .
\label{eq:WMC A}
\end{equation}
Through this condition
the temperature dependences of the quark and gluonic condensates
determine the intrinsic temperature dependence
of the bare parameter $F_\pi(\Lambda;T)$,
which is then converted into 
those of the on-shell parameter $F_\pi(\mu=0;T)$ 
through the Wilsonian RGE.

Now, let us consider the Wilsonian matching near the
chiral symmetry restoration point
assuming that the quark condensate approaches zero
continuously for $T \to T_c$.\footnote{
  Here and henceforth, I use just $T_c$ 
  which actually implies $T_c - \epsilon$.
}
First, note that
the Wilsonian matching condition~(\ref{eq:WMC A}) 
provides
\begin{equation}
  \frac{F^2_\pi (\Lambda ;T_c)}{{\Lambda}^2} 
  = \frac{1}{8{\pi}^2}\Bigl[
                            1 + \frac{\alpha _s}{\pi} +
                             \frac{2{\pi}^2}{3}
                            \frac{\langle \frac{\alpha _s}{\pi}
                            G_{\mu \nu}G^{\mu \nu} \rangle_{T_c} }
                             {{\Lambda}^4}
                 \Bigr]
 \neq 0 
\ ,
\label{eq:WMC A Tc}
\end{equation}
which implies that the matching with QCD dictates
\begin{equation}
F^2_\pi (\Lambda ;T_c) \neq 0 
\label{Fp2 Lam Tc}
\end{equation}
even at $T_c$ where the on-shell $\pi$ decay
constant approaches zero
by adding the quantum corrections through
the RGE including the quadratic divergence~\cite{HY:WM}
and the hadronic thermal correction~\cite{HS:VMT,HS:VD}.
As was shown in Ref.~\cite{HKR} for the VM in dense matter,
Lorentz non-invariant version of
the VM conditions for the bare parameters are obtained 
by the requirement of the equality between the axial-vector
and vector current correlators in the HLS,
which should be valid also in hot matter~\cite{HKRS}:
\begin{eqnarray}
 && a_{\rm{bare}}^t \equiv
  \Biggl( \frac{F_{\sigma,\rm{bare}}^t}{F_{\pi,\rm{bare}}^t} \Biggr)^2
  \ \mathop{\longrightarrow}_{T \rightarrow T_c}\ 
  1, \quad
  a_{\rm{bare}}^s \equiv
  \Biggl( \frac{F_{\sigma,\rm{bare}}^s}{F_{\pi,\rm{bare}}^s} \Biggr)^2
  \ \mathop{\longrightarrow}_{T \rightarrow T_c}\ 
  1, \label{EVM a}\\
 && g_{T,\rm{bare}} 
    \ \mathop{\longrightarrow}_{T \rightarrow T_c}\ 
    0, \quad
    g_{L,\rm{bare}}
    \ \mathop{\longrightarrow}_{T \rightarrow T_c}\ 
    0, \label{EVM g}
\end{eqnarray}
where $a^t_{\rm{bare}}, a^s_{\rm{bare}}, g_{T,\rm{bare}}$ and 
$g_{L,\rm{bare}}$ are the extensions of the parameters
$a_{\rm{bare}}$ and $g_{\rm{bare}}$ in the bare Lagrangian
with the Lorentz symmetry breaking effect included as in Appendix A
of Ref.~\cite{HKR}.

When we use the conditions for the parameters $a^{t,s}$ 
in Eq.~(\ref{EVM a})
and the above result that the Lorentz symmetry violation 
between the bare $\pi$ decay constants 
$F_{\pi,\rm{bare}}^{t,s}$ is small, 
we can easily show
that the Lorentz symmetry breaking effect between
the temporal and spatial bare $\sigma$ decay constants is also small,
$F_{\sigma,\rm{bare}}^t \simeq F_{\sigma,\rm{bare}}^s$~\cite{HKRS}.
While we cannot determine the ratio $g_{L,\rm{bare}}/g_{T,\rm{bare}}$
through the Wilsonian matching
since the transverse mode of $\rho$ decouples near $T_c$.
However, this implies that the transverse mode is irrelevant
to the quantities studied in the present analysis.
Therefore, I set
$g_{L,\rm{bare}}=g_{T,\rm{bare}}$ for simplicity and
use the Lorentz invariant Lagrangian at bare level.
In the low temperature region, the intrinsic temperature dependences
are negligible, so that one can use the 
Lorentz invariant Lagrangian at bare level
as in the analysis by the ordinary chiral
Lagrangian in Ref.~\cite{GLT}.

As I discussed above, in a good approximation,
one can start from the Lorentz invariant
bare Lagrangian even in hot matter.
In such a case
the axial-vector and the vector
current correlators $G_A^{\rm{(HLS)}}$ and $G_V^{\rm{(HLS)}}$
are expressed by the same forms
as those at zero temperature with the bare parameters
having the intrinsic temperature dependences~\cite{HS:VMT}:
\begin{eqnarray}
 G^{\rm{(HLS)}}_A (Q^2) 
  &=& \frac{F^2_\pi (\Lambda;T)}{Q^2} -
      2z_2(\Lambda;T), \nonumber\\
 G^{\rm{(HLS)}}_V (Q^2) 
  &=& \frac{F^2_\sigma (\Lambda;T)
         [1 - 2g^2(\Lambda;T)z_3(\Lambda;T)]}
           {{M_\rho}^2(\Lambda;T) + Q^2} - 2z_1(\Lambda;T)\ .
  \label{correlator HLS at zero-T}
  \end{eqnarray}

When the critical temperature is approached from below up to
$T_c$,
the axial-vector and vector current correlators
derived in the OPE approach each other
for any value of $Q^2$.
Thus we require that
these current correlators in the HLS become close to each other
at $T_c$
for any value of $Q^2\ \mbox{around}\ {\Lambda}^2$.
By taking account of the fact 
$F^2_\pi (\Lambda ;T_c) \neq 0$ derived from
the Wilsonian matching condition 
given in Eq.~(\ref{eq:WMC A Tc}),
the requirement 
$\displaystyle
G_A^{(\rm{HLS})} \ \mathop{\longrightarrow}_{T \rightarrow T_c}\ 
G_V^{(\rm{HLS})}$ is satisfied
only if the following conditions are met~\cite{HS:VMT}: 
\begin{eqnarray}
&&
g(\Lambda;T) \mathop{\longrightarrow}_{T \rightarrow T_c} 0 \ ,
\qquad
a(\Lambda;T) \mathop{\longrightarrow}_{T \rightarrow T_c} 1 \ ,
\nonumber\\
&&
z_1(\Lambda;T) - z_2(\Lambda;T) 
\mathop{\longrightarrow}_{T \rightarrow T_c} 0 \ .
\label{g a z12:VMT}
\end{eqnarray}
These conditions (``VM conditions in hot matter'')
for the bare parameters
are converted into the
conditions for the on-shell parameters through the Wilsonian RGEs.
Since $g=0$ and $a=1$ are separately the fixed points of the RGEs for
$g$ and $a$~\cite{HY:letter},
the on-shell parameters also satisfy
$(g,a)=(0,1)$, and thus the parametric $\rho$ mass
satisfies $M_\rho = 0$.

Now, let me
include the hadronic thermal effects to obtain the $\rho$ pole
mass near $T_c$.
As I explained above,
the intrinsic temperature dependences imply that
$M_\rho/T \rightarrow 0$
for $T \rightarrow T_c$,
so that the $\rho$ pole mass near the
critical temperature is expressed as~\cite{HS:VMT,HS:VD}
\begin{eqnarray}
&& m_\rho^2(T)
  = M_\rho^2 +
  g^2 N_f \, \frac{15 - a^2}{144} \,T^2
\ .
\label{mrho at T 2}
\end{eqnarray}
Since $a \rightarrow 1$ near $T_c$,
the second term is positive. 
Then the $\rho$ pole mass $m_\rho$
is bigger than the parametric
$M_\rho$ due to the hadronic thermal corrections.
Nevertheless, 
{\it the intrinsic temperature dependence determined by the
Wilsonian matching requires
that the $\rho$ becomes massless at the
critical temperature}:
\begin{eqnarray}
&&
m_\rho^2(T)
\rightarrow 0 \ \ \mbox{for} \ T \rightarrow T_c \ ,
\end{eqnarray}
since the first term in Eq.~(\ref{mrho at T 2})
vanishes as $M_\rho\rightarrow 0$, and the second
term also vanishes since $g\rightarrow 0$ for $T \rightarrow T_c$.
This implies that
{\it the vector manifestation (VM) actually
occurs at the critical
temperature}~\cite{HS:VMT}.

\section{Summary}
\label{sec:sum}

In this write-up,
I first explained how the 
systematic chiral perturbation in the hidden local symmetry (HLS)
is justified in the large $N_c$ limit of QCD in section~\ref{sec:HLS}.
Next in section~\ref{sec:WM},
I summarized the basic concept of the Wilsonian matching,
through which some of the bare parameters of the HLS are 
determined by matching the HLS to the operator product
expansion in QCD.
In section~\ref{sec:VM} 
I showed the difference between the 
VM and
the conventional manifestation of chiral symmetry restoration
based on the linear sigma model
in terms of the chiral representation of the mesons.
Then, in section~\ref{sec:VMT}, I
reviewed how to formulate the VM in hot matter.

There are several predictions of the VM in hot matter
made so far.
In Ref.~\cite{HKRS},
the vector and axial-vector susceptibilities were studied.
It was shown that the equality between two susceptibilities
are satisfied and that the VM predicts
$\chi_A = \chi_V = \frac{2}{3} \, T_c^2$ for $N_f = 2$,
which is in good agreement with the result obtained in the lattice 
simulation~\cite{Allton}.
In Ref.~\cite{HS:VD},
a prediction associated with the validity of 
vector dominance (VD) in hot matter was made:
As a consequence of including the intrinsic effect,
the VD is largely violated at the critical temperature.
In addition to the above predictions,
the pion velocity was studied
including the effect of Lorentz symmetry 
breaking~\cite{Sasaki:Vpi,HKRS:pvT}.
It was shown that the pion velocity near $T_c$ is close to the speed
of light.
Furthermore, in Ref.~\cite{HRS:D},
starting with an HLS Lagrangian at the 
VM fixed point that incorporates the heavy-quark
symmetry and matching the bare theory to QCD, we calculated the
splitting of chiral doublers of $D$ mesons proposed in
Refs.~\cite{Nowak-Rho-Zahed:93,Bardeen-Hill:94,Nowak-Rho-Zahed:03},
and showed that the splitting comes out to be $0.31\pm0.12$\,GeV,
which is in good agreement with the 
experiment~\cite{BABAR,CLEO,Belle}.
Furthermore,
the matching showed that the mass splitting is 
directly proportional to the
light-quark condensate $\langle\bar{q}q\rangle$, 
which implies that
the splitting vanishes at the chiral
restoration point.

\section*{Acknowledgment}

I would like to thank Doctor Toshiki Fujimori,
Doctor Youngman Kim, Professor Mannque Rho,
Doctor Chihiro Sasaki and
Professor Koichi Yamawaki for collaboration in
several works done for the chiral perturbation
theory with hidden local symmetry and the vector manifestaion
on which this talk is based.
This work is supported in part by
the JSPS Grant-in-Aid for Scientific Research (c) (2) 16540241,
and by
the 21st Century COE
Program of Nagoya University provided by Japan Society for the
Promotion of Science (15COEG01).

\end{document}